%

\magnification 1000

\hsize= 16 truecm
\vsize= 22 truecm

\hoffset=-0.0 truecm
\voffset= +1 truecm

\baselineskip=14 pt

\parindent 15pt 

\footline={\iftitlepage{\hfil}\else
       \hss\tenrm-- \folio\ --\hss\fi}	

\parskip 0pt 


\font\bello= cmr10 scaled \magstep2
\font\piccolo=cmr8

\font\pbf=cmbx8

\def\eq{\autoeqno}
\def\re{\eqrefp}


\def\eq#1{\autoeqno{#1}}
\def\re#1{\eqrefp{#1}}

\newcount\notenumber \notenumber=1
\def\nota#1{\unskip\footnote{$^{\the\notenumber}$}{\piccolo #1}%
  \global\advance\notenumber by 1}

\def\s{\scriptstyle}  

\catcode`@=11 
%

\newcount\cit@num\global\cit@num=0

\newwrite\file@bibliografia
\newif\if@bibliografia
\@bibliografiafalse

\def\lp@cite{[}
\def\rp@cite{]}
\def\trap@cite#1{\lp@cite #1\rp@cite}
\def\lp@bibl{[}
\def\rp@bibl{]}
\def\trap@bibl#1{\lp@bibl #1\rp@bibl}

\def\refe@renza#1{\if@bibliografia\immediate        
    \write\file@bibliografia{
    \string\item{\trap@bibl{\cref{#1}}}\string
    \bibl@ref{#1}\string\bibl@skip}\fi}

\def\ref@ridefinita#1{\if@bibliografia\immediate\write\file@bibliografia{ 
    \string\item{?? \trap@bibl{\cref{#1}}} ??? tentativo di ridefinire la 
      citazione #1 !!! \string\bibl@skip}\fi}

\def\bibl@ref#1{\se@indefinito{@ref@#1}\immediate
    \write16{ ??? biblitem #1 indefinito !!!}\expandafter\xdef
    \csname @ref@#1\endcsname{ ??}\fi\csname @ref@#1\endcsname}

\def\c@label#1{\global\advance\cit@num by 1\xdef            
   \la@citazione{\the\cit@num}\expandafter
   \xdef\csname @c@#1\endcsname{\la@citazione}}

\def\bibl@skip{\vskip 0truept}


\def\stileincite#1#2{\global\def\lp@cite{#1}\global
    \def\rp@cite{#2}}
\def\stileinbibl#1#2{\global\def\lp@bibl{#1}\global
    \def\rp@bibl{#2}}

\def\citpreset#1{\global\cit@num=#1
    \immediate\write16{ !!! cit-preset = #1 }    }

\def\autobibliografia{\global\@bibliografiatrue\immediate
    \write16{ !!! Genera il file \jobname.BIB}\immediate
    \openout\file@bibliografia=\jobname.bib}

\def\cref#1{\se@indefinito                  
   {@c@#1}\c@label{#1}\refe@renza{#1}\fi\csname @c@#1\endcsname}

\def\cite#1{\trap@cite{\cref{#1}}}                  
\def\ccite#1#2{\trap@cite{\cref{#1},\cref{#2}}}     
\def\cccite#1#2#3{\trap@cite{\cref{#1},\cref{#2},\cref{#3}}}
\def\ccccite#1#2#3#4{\trap@cite{\cref{#1},\cref{#2},\cref{#3},\cref{#4}}}
\def\ncite#1#2{\trap@cite{\cref{#1}--\cref{#2}}}    
\def\upcite#1{$^{\,\trap@cite{\cref{#1}}}$}               
\def\upccite#1#2{$^{\,\trap@cite{\cref{#1},\cref{#2}}}$}  
\def\upncite#1#2{$^{\,\trap@cite{\cref{#1}-\cref{#2}}}$}  

\def\clabel#1{\se@indefinito{@c@#1}\c@label           
    {#1}\refe@renza{#1}\else\c@label{#1}\ref@ridefinita{#1}\fi}

\def\biblskip#1{\def\bibl@skip{\vskip #1}}           

\def\insertbibliografia{\if@bibliografia             
    \immediate\write\file@bibliografia{ }
    \immediate\closeout\file@bibliografia
    \catcode`@=11\input\jobname.bib\catcode`@=12\fi}


\def\commento#1{\relax} 
\def\biblitem#1#2\par{\expandafter\xdef\csname @ref@#1\endcsname{#2}}



%
%
\def\b@lank{ }


\newif\iftitlepage      \titlepagetrue

\def\titoli#1{
         \xdef\prima@riga{#1}\voffset+20pt
        \headline={\ifnum\pageno=1
             {\hfil}\else\hfil{\piccolo \prima@riga}\hfil\fi}}

\def\duetitoli#1#2{
                    \voffset=+20pt
                    \headline={\iftitlepage{\hfil}\else
                              {\ifodd\pageno\hfil{\piccolo #2}\hfil
             \else\hfil{\piccolo #1}\hfil\fi}\fi} }

\def\la@sezionecorrente{0}

\catcode`@=12


\autobibliografia

%
\catcode`@=11 
%
%
\def\b@lank{ }

\newif\if@simboli
\newif\if@riferimenti
\newif\if@bozze

\newwrite\file@simboli
\def\simboli{
    \immediate\write16{ !!! Genera il file \jobname.SMB }
    \@simbolitrue\immediate\openout\file@simboli=\jobname.smb}

\def\bozze{\@bozzetrue}

\newcount\eq@num\global\eq@num=0
\newcount\sect@num\global\sect@num=0

\newif\if@ndoppia
\def\numerazionedoppia{\@ndoppiatrue\gdef\la@sezionecorrente{\the\sect@num}}

\def\se@indefinito#1{\expandafter\ifx\csname#1\endcsname\relax}
\def\spo@glia#1>{} 

\newif\if@primasezione
\@primasezionetrue

\def\s@ection#1\par{\immediate
    \write16{#1}\if@primasezione\global\@primasezionefalse\else\goodbreak
    \vskip\spaziosoprasez\fi\noindent
    {\bf#1}\nobreak\vskip\spaziosottosez\nobreak\noindent}
%

\def\sezpreset#1{\global\sect@num=#1
    \immediate\write16{ !!! sez-preset = #1 }   }

\def\spaziosoprasez{26pt plus5pt minus3pt}
\def\spaziosottosez{15pt}

\def\sref#1{\se@indefinito{@s@#1}\immediate\write16{ ??? \string\sref{#1}
    non definita !!!}
    \expandafter\xdef\csname @s@#1\endcsname{??}\fi\csname @s@#1\endcsname}

\def\autosez#1#2\par{
    \global\advance\sect@num by 1\if@ndoppia\global\eq@num=0\fi
    \xdef\la@sezionecorrente{\the\sect@num}
    \def\usa@getta{1}\se@indefinito{@s@#1}\def\usa@getta{2}\fi
    \expandafter\ifx\csname @s@#1\endcsname\la@sezionecorrente\def
    \usa@getta{2}\fi
    \ifodd\usa@getta\immediate\write16
      { ??? possibili riferimenti errati a \string\sref{#1} !!!}\fi
    \expandafter\xdef\csname @s@#1\endcsname{\la@sezionecorrente}
    \immediate\write16{\la@sezionecorrente. #2}
    \if@simboli
      \immediate\write\file@simboli{ }\immediate\write\file@simboli{ }
      \immediate\write\file@simboli{  Sezione 
                                  \la@sezionecorrente :   sref.   #1}
      \immediate\write\file@simboli{ } \fi
    \if@riferimenti
      \immediate\write\file@ausiliario{\string\expandafter\string\edef
      \string\csname\b@lank @s@#1\string\endcsname{\la@sezionecorrente}}\fi
    \goodbreak\vskip 24pt plus 60pt
    \noindent\if@bozze\llap{\it#1\quad }\fi
      {\bf\the\sect@num.\quad #2}\par\nobreak\vskip 15pt
    \nobreak\noindent}

\def\semiautosez#1#2\par{
    \gdef\la@sezionecorrente{#1}\if@ndoppia\global\eq@num=0\fi
    \if@simboli
      \immediate\write\file@simboli{ }\immediate\write\file@simboli{ }
      \immediate\write\file@simboli{  Sezione ** : sref.
          \expandafter\spo@glia\meaning\la@sezionecorrente}
      \immediate\write\file@simboli{ }\fi
    \s@ection#2\par}


\def\eqpreset#1{\global\eq@num=#1
     \immediate\write16{ !!! eq-preset = #1 }     }

\def\eqref#1{\se@indefinito{@eq@#1}
    \immediate\write16{ ??? \string\eqref{#1} non definita !!!}
    \expandafter\xdef\csname @eq@#1\endcsname{??}
    \fi\csname @eq@#1\endcsname}

\def\eqlabel#1{\global\advance\eq@num by 1
    \if@ndoppia\xdef\il@numero{\la@sezionecorrente.\the\eq@num}
       \else\xdef\il@numero{\the\eq@num}\fi
    \def\usa@getta{1}\se@indefinito{@eq@#1}\def\usa@getta{2}\fi
    \expandafter\ifx\csname @eq@#1\endcsname\il@numero\def\usa@getta{2}\fi
    \ifodd\usa@getta\immediate\write16
       { ??? possibili riferimenti errati a \string\eqref{#1} !!!}\fi
    \expandafter\xdef\csname @eq@#1\endcsname{\il@numero}
    \if@ndoppia
       \def\usa@getta{\expandafter\spo@glia\meaning
       \la@sezionecorrente.\the\eq@num}
       \else\def\usa@getta{\the\eq@num}\fi
    \if@simboli
       \immediate\write\file@simboli{  Equazione 
            \usa@getta :  eqref.   #1}\fi
    \if@riferimenti
       \immediate\write\file@ausiliario{\string\expandafter\string\edef
       \string\csname\b@lank @eq@#1\string\endcsname{\usa@getta}}\fi}

\def\autoreqno#1{\eqlabel{#1}\eqno(\csname @eq@#1\endcsname)
       \if@bozze\rlap{\it\quad #1}\fi}
\def\autoleqno#1{\eqlabel{#1}\leqno\if@bozze\llap{\it#1\quad}
       \fi(\csname @eq@#1\endcsname)}
\def\eqrefp#1{(\eqref{#1})}
\def\numeriadestra{\let\autoeqno=\autoreqno}
\def\numeriasinistra{\let\autoeqno=\autoleqno}
\numeriadestra

\catcode`@=12

\numerazionedoppia


\def\bbuildrel#1_#2{\mathrel{\mathop{\kern 0pt#1}\limits_{#2}}}


\vglue 3 truecm
{\parindent 0 pt

{\bello On the stationary points of the TAP free energy.}

\vskip 1 truecm
Andrea Cavagna, Irene Giardina and Giorgio Parisi
\vskip 0.5 truecm
{\it          Dipartimento di Fisica,
Universit\`a di Roma I `La Sapienza',
P.le A. Moro 5, 00185 Roma, Italy 

INFN Sezione di Roma I, Roma, Italy

\vskip 0.5 truecm                  }

{\sl andrea.cavagna@roma1.infn.it}

{\sl irene.giardina@roma1.infn.it}

{\sl giorgio.parisi@roma1.infn.it}

\vskip 1 truecm 

October 1997       }
 
\vskip 2 truecm 

\noindent
{\bf Abstract.}
\vskip 0.5 truecm

\noindent
In the context of the $p$-spin spherical model, we introduce a method for 
the computation of the number of stationary points of any nature (minima, 
saddles, etc.) of the TAP free energy. In doing this we clarify the ambiguities
related to the approximations usually adopted in the standard calculations 
of the number of states in mean field spin glass models. 


\vskip 1 truecm


\vfill\eject

\titlepagefalse

\duetitoli{\it A Cavagna, I Giardina and G Parisi}{\it On the stationary 
points of the TAP free energy}

\autosez{intro} Introduction.
\par

\noindent
Mean field spin glass models are characterized in their low temperature phase
by the great number of metastable as well as equilibrium states. 
A question which naturally arises in this context is the computation
of the number ${\cal N}$ of these states, or,
more precisely, the analysis of how this number increases 
with the size $N$ of the system.

In models with a continuous transition, as the SK model \cite{sk},
the equilibrium thermodynamics is dominated by a number of states that 
remains finite when $N\to\infty$, while there is an 
exponentially high number of metastable states \cite{braymoore}, 
which do not contribute to the thermodynamics of the system. 
On the other hand, models with a discontinuous transition,
as the $p$-spin spherical model \cite{grome}\cite{tirumma}\cite{crisa1}, 
exhibit a temperature range where the number
of metastable {\it and} equilibrium states with given 
energy density $E$ grows exponentially, i.e  
${\cal N}(E)\sim \exp(N\Sigma(E))$ \cite{crisatap}.
In this last case the knowledge of the {\it complexity} $\Sigma(E)$ 
is crucial, since it gives a finite entropic contribution to the global free 
energy \cite{kpz}.
It is therefore particularly important in this case to have a 
well defined method to compute the number of states of the system.

The standard strategy to perform this calculation is grounded on the 
formulation of mean field equations for the local magnetizations, 
the TAP equations \cite{tap}. 
The solutions of these equations are identified with 
equilibrium or metastable states of the system and therefore one simply 
resorts to count the number of these solutions. 

This standard approach contains however some ambiguities.
The TAP solutions can be viewed 
as the stationary points of a TAP free energy $f_{TAP}$, function of the 
magnetizations \cite{braymoore}\cite{crisatap}\cite{kpz}, 
therefore only the minima of this 
free energy can actually be identified with metastable or equilibrium states
of the system. Yet, there are surely many other kinds of stationary
points different from minima. 
When in the standard approach one counts the number of 
TAP solutions it is not clear whether only the genuine states of the system  
are taken into consideration.

Moreover, a typical approximation of the standard method is 
related to the modulus of the determinant of the free energy Hessian 
(i.e. the Jacobian of the equations), which appears 
in the integral over all the solutions \cite{kurchan}. 
The presence of this  modulus is fundamental to avoid a trivial result: if 
one tries to count the number of stationary points of a function {\it without}
this modulus, each stationary point is weighted with the sign of
the Hessian and one obtains a simple topological constant, by virtue of the
Morse theorem \cite{morse}. 
Nonetheless, in the standard approach this modulus is always
disregarded to simplify the computation.

From what said above we are leaded to say that the standard procedure 
is not really under control. Nonetheless, at least in the case of the $p$-spin
spherical model, this standard calculation gives a result \cite{crisatap},
that has been
exactly confirmed by a completely different approach \cite{franzparisi}. 
This result is therefore
correct, although all the approximations involved are not well justified.
On the other hand for the case of the SK model there is no confirmation of
the standard result of \cite{braymoore}.

The aim of this paper is to clarify this subject, at least in 
the case of the $p$-spin spherical model.
In the context of the replica approach we show that different solutions of the
saddle point equations for the overlap matrix are related to different 
kinds of stationary points (minima, saddles, etc.). Grouping them into 
classes characterized by the 
number $k$ of their instable directions, we find that each class 
has a different complexity $\Sigma_k(E)$. By virtue of this result 
we are able to extract 
separately from the total number of solutions the contribution of minima and
of saddles of various indices $k$, discovering that there is an ordering 
of the complexities $\Sigma_k(E)$: 
at a given energy $E$ only {\it one} kind of 
stationary points is exponentially dominant over all the others, so that in 
the thermodynamic limit the weight of the sign of the determinant
has no influence. Therefore, as long as the energy is kept fixed,  
the modulus can be disregarded and the standard approach gives the correct 
result.


\autosez{ilcalcolo} The complexity.
\par
\noindent
The $p$-spin spherical model is defined by the Hamiltonian
$$
H(s)=-\!\!\!\!\sum_{i_1<\dots <i_p} J_{i_1\dots i_p} s_{i_1}\dots s_{i_p}
\ .
\eq{mod}
$$
The spins $s$ are real variables satisfying the spherical
constraint $\sum_i s_i^2 = N$, where $N$ is the size of the system. The
couplings  $J$ are Gaussian variables with zero mean and variance
$p!/2N^{p-1}$. 
In the context of the TAP approach \cite{tap}, one formulates a set of 
mean field
equations for the local magnetizations $m_i=\langle s_i\rangle$.  
In \cite{kpz} it has been introduced a free energy density
$f_{TAP}$, function of the magnetizations $m_i$. The minimization of $f_{TAP}$
with respect to $m_i$ gives the TAP equations of the system.
We can express the magnetization vector $m$  in terms of its angular 
part $\sigma$ and of its self-overlap $q=1/N \sum_i {m_i}^2$:
$$
m_i =  \sqrt{q}\ \sigma_i \quad ;
\quad \quad \sigma \cdot \sigma = \sum_i \sigma_i^2 = N \ .
\eq{tra}
$$
The TAP equations for $\sigma$ read \cite{kpz}
$$
0 = - p\!\!\sum_{i_2<\dots <i_p}J_{l,i_2\dots i_p}\,
\sigma_{i_2}\dots\sigma_{i_p} - pE\sigma_l
\ \buildrel\hbox{\sevenrm def}\over= {\cal T}_l(\sigma;E) 
\quad,\quad l=1,\dots,N
\eq{tap}
$$
where $E$ is the zero temperature energy density
$$
E=-{1\over N}\!\sum_{i_1<\dots<i_p}J_{i_1\dots i_p}
\sigma_{i_1}\dots \sigma_{i_p}  \ .
\eq{ene}
$$
In the following we shall always refer to the zero temperature 
energy density.
The equations for $\sigma$ do not depend on the temperature,  
while the equation for $q$ does \cite{kpz}. 
Moreover the $q$ equation has solution as long as the energy density is
lower than a maximum value of the energy, called {\it threshold } 
energy $E_{th}$. The dependence on 
temperature of the set of TAP solutions $\{m(T)\}_{\alpha=1\dots{\cal N}}$ 
comes entirely from $q$, 
while their multiplicity ${\cal N}$ is encoded in equations \re{tap} and thus 
does not depend on the temperature. 
It turns out that there is an exponentially high number of solutions of \re{tap} 
for each given value of the energy density $E$, 
${\cal N}(E) \sim \exp(N \Sigma(E))$, where $\Sigma(E)$
is the complexity, computed for this model in \cite{crisatap}.
$\Sigma(E)$ is an increasing function of $E$, which reaches a finite value
for $E=E_{th}$. To avoid any confusion, we note that the TAP free energy density 
of a solution at temperature $T$ is unambiguously determined by its zero
temperature energy density $E$. Therefore in the following we shall always
use $E$ to label TAP solutions.

We start our analysis with the computation of $\Sigma(E)$, paying special 
attention to the nature of the stationary points actually considered.
By definition we write
$$
\Sigma(E)
\ \buildrel\hbox{\sevenrm def}\over=
\ \lim_{N\to \infty}{1\over N} \ 
\overline{   \log {\cal N}(E) } 
\eq{lei}
$$ 
We average the logarithm of $\cal N$ since this 
is the extensive quantity. To perform this average it is
necessary to introduce replicas already at this level of the calculation. 
However, it can be shown that 
the correct ansatz for the overlap matrix is symmetric and 
diagonal and this is equivalent to average directly the number $\cal
N$ of  the solutions. Therefore we will
perform the annealed computation:
$$
\Sigma(E)
\ =
\ \lim_{N\to \infty}{1\over N}
\log \overline{  {\cal N}(E) } \ . 
\eq{lei2}
$$ 
In terms of the angular parts \re{tra} we have
$$
\Sigma (E)  
= \lim_{N \to \infty}{1\over N} \log \overline{
\int  {\cal  D}\sigma  \ \delta(\sigma \cdot \sigma -N)   
\prod_{l=1}^N\ \delta ({\cal T}_l(\sigma;E)) \  
\left| \det {\cal H}(\sigma;E) \right| }
\eq{mink}
$$
where  ${\cal H}(\sigma;E)$ is the Hessian of the TAP equations evaluated  in 
the solution $\sigma$ of energy density $E$. It is given by
$$	
{\cal H}_{r,l}(\sigma;E)=
{\partial {\cal T}_r(\sigma;E) \over \partial 
\sigma_l}=
- p(p-1) \!\!\sum_{i_3< \dots <i_p}J_{r,l,i_3\dots i_p}\ 
\sigma_{i_3}\dots\sigma_{i_p} - p E \delta_{r,l} \ .
\eq{hess}
$$
We stress that by means of formula \re{mink} we are counting only 
the solutions with a {\it given} energy density $E$. 
This is a crucial point: the principal effort of our discussion will be to 
show that, as long as $E<E_{th}$, if we keep the energy fixed 
the modulus in \re{mink} can be dropped without affecting 
the result in the limit $N\to\infty$. We shall return on this point with
greater details at the end of our discussion. 
We therefore perform the calculation without the modulus,
showing {\it a posteriori} which are the justifications of this 
procedure.
Let us introduce a Bosonic representation both for the determinant and 
the delta functions that implement the TAP equations:
$$
\eqalign{
\det {\cal H}=
\ \lim_{n \to -2} \left \{ \det {\cal H} \right \}^{-{n \over 2}}= &
\ \lim_{n\to -2} 
\int {\cal D} \phi^a 
\exp{  \left ( -{1 \over 2} \sum_{a=1}^n (\phi^a {\cal H} \phi^a) \right )   }
 \cr  
\prod_{l=1}^N\ \delta ({\cal T}_l(\sigma;E))=& \ 
\int {\cal D} \mu \exp 
(i\mu{\cal T}) \cr}
\eq{del}
$$
where the sums over repeated site indices are understood.
The average over the disorder generates couplings between the  fields 
$\phi$, $\sigma$ and $\mu$. A crucial approximation is to set equal to zero 
the couplings $\phi^a\cdot\sigma$ and $\phi^a\cdot\mu$ which depend on one 
replica index and which break the rotational invariance in 
the space of the replicas. We will see that this approximation is consistent 
with all the solutions we shall consider for the saddle point equations.
Thus we retain only the terms $\phi_a\cdot\phi_b$,
$\mu\cdot\mu$ and $\mu\cdot\sigma$.
It is easy to see that this approximation is equivalent to write
$$
\Sigma (E)  
= \lim_{N\to\infty} {1\over N} \log 
\int  {\cal  D}\sigma \   \delta(\sigma \cdot \sigma -N) \ \ 
\overline   {  \prod_{l=1}^N\ \delta({\cal T}_l(\sigma;E))   }  \ 
\times \ 
\overline { \det{\cal H}(\sigma;E) }
 \  .
\eq{fatt}
$$
Once
averaged over the disorder, because of the spherical constraint the part
of the determinant does not depend on $\sigma$ any more. Therefore we have
$$
\Sigma (E)  =  A(E) \ + \ B(E) 
\eq{bada}
$$
$$
\eqalign{
& A(E) = \lim_{N \to \infty}{1\over N} \log 
\int  {\cal  D}\sigma \   \delta(\sigma \cdot \sigma -N) \ \  
\overline{  
\int {\cal D} \mu \exp 
(i\mu{\cal T})
} \cr
& B(E) =  \lim_{N \to \infty} {1\over N} \log 
\int  {\cal  D}\sigma \   \delta(\sigma \cdot \sigma -N) \ \ 
\overline {   
\lim_{n\to -2} 
\int {\cal D} \phi^a 
\exp\left ( -{1 \over 2} \sum_{a=1}^n (\phi^a {\cal H} \phi^a) \right )
}\  .  \cr}
\eq{fatt2}
$$
The first integral does not involve replicas and gives the following 
contribution
$$
A(E)= {1\over 2} -  {1\over 2} \log{{p\over 2}} -E^2  \ .
\eq{A} 
$$ 
The second integral is  more subtle to solve because it contains replicas
and an appropriate ansatz has to be chosen to solve the saddle point equations. 
Moreover this integral is the one related to the Hessian of the TAP solutions 
and thus it contains information on the nature of the solutions 
(minima, saddles or maxima) that we are counting.
Once averaged over the disorder and introduced the overlap matrix 
$Q_{ab}=-p(p-1)(\phi^a\cdot \phi^b)/2 N$, we obtain
$$
B(E)=\lim_{N\to\infty} {1\over N} \log  \ \lim_{n \to -2} 
\int {\cal D}Q_{ab} \exp \left \{ -N \  
\left ( { {\rm Tr}\ Q^2 \over {2p(p-1)} } + {1\over 2}
\log\det(-pE+Q) \right ) \right \}
 \ .
\eq{Ba}
$$
As an ansatz for the matrix $Q$ we take $Q_{ab}=q_a\ \delta_{ab}$.
In this way the exponent of \re{Ba} splits into $n$ independent parts,
each one giving the same saddle point equation for $q_a$, whose possible
solutions are
$$
q_a=q_{\pm}={p\over 2}\left(E \pm \sqrt{E^2-E_{th}^2}\right)
 \ \ , \ \ E_{th}=-\sqrt{2 (p-1) \over p}  \ .
\eq{sol}
$$
We restrict our discussion to $E<E_{th}$, so that $q_\pm$ are real.
Since each $q_a$ can assume one of these two values, we
have a multiplicity of different solutions.
The analysis of the fluctuations shows that there is a stable solution 
${\cal S}_0$, given by 
$$
{\cal S}_0: \quad\quad\quad q_a=q_{+} \ \ ,\ \ a=1,\dots, n
\eq{s0}
$$ 
This solution is invariant under rotations in the replica space and thus 
the approximation we made setting to zero the terms 
depending on one replica index turns out to be consistent.
The solution ${\cal S}_0$ gives the complexity 
$$
\Sigma_0(E)={ q_{+}^2\over p(p-1)}+
\log(-pE+q_+)
+A(E)
\eq{solB}
$$
with $A$ given in \re{A}. This is the known result 
of \cite{crisatap}. It is important to note that 
this result has been 
confirmed in the analysis of \cite{franzparisi} and \cite{noi} where, 
by means of a completely different method, 
it has been shown that $\Sigma_0$
is equal to the logarithm of the number of genuine states of the 
system, and thus that $\Sigma_0$ is the complexity of the {\it minima} of 
the TAP free energy.
 
Nonetheless, we note the presence of many other solutions of the saddle
point equations, involving
both the values $q_{\pm}$. In particular we are interested in 
the solution ${\cal S}_1$ with the lowest degree of instability, that is
$$
{\cal S}_1:\quad\quad\quad q_1=q_{-} \ \ , 
\ \ q_a=q_{+} \ \ ,\ \ 
a=2,\dots,n
\eq{nana}
$$
(and permutations). 
This solution presents a one step breaking of the rotational invariance 
in the replica space.
Therefore one can concern about the fact that we have disregarded 
terms breaking this invariance.
To check this point, we have performed the whole computation retaining the 
terms $\phi^a\cdot\sigma$, $\phi^a\cdot\mu$ and we have looked for a 
solution breaking the rotational invariance in the replica space. We found
analytically 
that the saddle point equations
give as a unique solution $\phi^a\cdot\sigma=0$ and $\phi^a\cdot\mu=0$ and 
thus that solution ${\cal S}_1$ is recovered.
The complexity $\Sigma_1$ arising from ${\cal S}_1$ is
$$
\Sigma_1(E)={3\over 2}{q_{+}^2\over p(p-1)}-{1\over 2}{q_{-}^2\over p(p-1)}
        +{3\over 2}\log(-pE+q_+)
        -{1\over 2}\log(-pE+q_-)
+A(E)
\eq{pedro}
$$
which is {\it lower} than $\Sigma_0$, since $|q_-|\geq |q_+|$. 
In this context
it is not clear which is the physical meaning of the complexity $\Sigma_1$,
neither if there is one. 
Moreover, apart from the fact that the complexity $\Sigma_0$
is confirmed by a different method to be related to the number of minima, 
we have given no justification of dropping the modulus in the 
original formula.
We shall see in the next sections that the analysis of the average spectrum of 
the TAP Hessian gives an answer to both these questions.


\autosez{spettro} The Hessian spectrum.
\par
\noindent
In the previous section we made the approximation 
of setting to zero the couplings  $\phi^a\cdot\sigma$ and $\phi^a\cdot\mu$.
We stress that this approximation is consistent when considering the solutions 
${\cal S}_0$ and  ${ \cal S}_1$.
As a consequence, what appears in expression \re{fatt}
is the Hessian function evaluated in a {\it generic} vector
$\sigma$, and not in a TAP solution. 
This means that, in the context on this approximation, 
the properties of the TAP Hessian that are relevant in 
determining the behaviour of $\Sigma$, are well encoded in the  matrix
${\cal H}(\sigma;E)$ which has the same functional form of the TAP Hessian, but
requires $\sigma$  only to satisfy the spherical constraint.
The average spectrum is then defined in the following way
$$
\rho(\lambda;E)= 
\lim_{N\to\infty}\overline{ 
 \int{\cal  D}\sigma \ \delta(\sigma \cdot \sigma -N)
                     \ \rho_J(\lambda;\sigma)}  
\eq{zum}
$$
where $\rho_J(\lambda;\sigma)$ is  the spectrum for a given realization
of the disorder whose expression is:
$$
\rho_J(\lambda;\sigma)= -{1\over N \pi}\  {\rm Im}\ {\rm Tr}
 ( {\cal H} - \lambda + i \epsilon  )^{-1}  \  .
\eq{spettj}
$$
We can write the trace in the following way:
$$
\eqalign{
&{\rm Tr} ( {\cal H} - \lambda + i \epsilon  )^{-1}=
\sum_{l=1}^{N} [( {\cal H} - \lambda + i \epsilon  )^{-1}]_{ll}=   \cr
=&\lim_{n \to 0} \ \int {\cal D}\phi^a\ \phi^1\cdot \phi^1
\ \exp \left \{ -{1 \over 2} \sum_{a=1}^n \phi^a 
({\cal H} - \lambda + i \epsilon)  \phi^a
\right \}   \  . \cr
}
\eq{traccia}
$$
Once averaged over the disorder $J$ and exploited the spherical constraint on
$\sigma$, this computation becomes   
analogous to the one of the average spectrum of a Gaussian ensemble of 
symmetric random matrices \cite{metha}. If  we introduce the overlap matrix 
$Q_{ab}=-p(p-1)(\phi_a\cdot\phi_b)/2N$ we finally get
$$
\eqalign{
\rho(\lambda;E)= \lim_{N\to\infty} -{1\over N\pi }\  
{\rm Im}\  \lim_{n \to 0}  
\int & {\cal D}Q_{ab}  \  (- pE-\lambda + Q)^{-1}_{11}\  \times  \cr 
& \times \ \exp \left \{ -N \  
\left ( { {\rm Tr}\  Q^2 \over 2p(p-1)} + {1\over 2}
\log\det (- pE-\lambda + Q) \right ) 
\right \}                                                       \cr
}
\eq{puppa}
$$
It is important to note the great similarity between equations \re{puppa} and
\re{Ba}. If we choose once again a diagonal ansatz  $Q_{ab}=w_a \delta_{ab}$, 
we get the following  solutions of the saddle point equations
$$
w_a=w_{\pm}(\lambda)={p\over 2} \left( {\lambda \over p} + E  \pm 
\sqrt{\left({\lambda \over p}  + E \right)^2- E_{th}^2  }  
\ \right)
\eq{solw}
$$
where $E_{th}$ is the same as in equation \re{sol}. For $\lambda=0$ the 
integrand in \re{puppa} is identical to the one of \re{Ba} and 
$w_{\pm}(0)=q_{\pm}$.
As in the case of the complexity we have a multiplicity of different solutions.
To get a finite contribution to $\rho$ it is necessary that the argument of 
the exponential in \re{puppa} is zero. Since $n\to 0$, this can be achieved
taking the same value for each $w_a$. Moreover, the condition
$\rho \geq 0$ shows that we must take the solution
$$
{\cal S}_0: \quad\quad\quad  w_a=w_{+}(\lambda) \ \ , \ \ a=1,\dots,n
\eq{s0ro}
$$
that is exactly the same kind of solution that leaded to $\Sigma_0$.  
If we look at \re{solw} we can see that the non-zero contribution to 
$\rho$ comes from the region $ -pE+pE_{th}  <\lambda < -pE- 
pE_{th}$,  where $w_{+}$ develops an imaginary part. Thus
$$  
\rho_0(\lambda;E) = {1\over \pi p(p-1)} \sqrt{ p^2 E_{th}^2-
\left(\lambda +pE\right)^2 }  \ .
\eq{semi} 
$$
We stress that the solution ${\cal S}_0$ is the only one that gives a finite
contribution $\rho_0$ to $\rho$. Formula \re{semi}
is the well known Wigner semicircle law \cite{wigner}, that can be 
obtained for symmetric
Gaussian random matrices also without using replicas \cite{metha}. 
This result tells us that for $E <E_{th}$ the averaged spectrum has
a strictly positive support and thus the typical determinant of the Hessian is 
positive, i.e. that the dominant part of TAP solutions with energy 
density $E < E_{th}$  are {\it minima}. On the other hand, when $E$ approaches 
$E_{th}$ the lowest eigenvalue $\lambda=p(E_{th}-E)$ goes to
zero. Therefore the typical solutions with  $E=E_{th}$ have some 
flat directions \cite{ck1}.

We understand now the reason why the complexity of \re{solB}
is related to the number of minima: the solution ${\cal S}_0$
of the saddle point equations
leading to $\Sigma_0$ is exactly the same as the one leading to
the eigenvalue distribution $\rho_0$, which has positive support.

The important thing is that in this context it is possible to give a precise 
physical interpretation of the solution ${\cal S}_1$ of \re{nana}: 
as we are going to show in the next section, ${\cal S}_1$ is related to 
the exponentially small corrections to the distribution $\rho_0$ 
and therefore gives informations on those TAP solutions which are not 
minima.    
 

\autosez{code} Exponential tails and complexity of the saddles.
\par 
\noindent
For an ensemble of symmetric random matrices with a Gaussian distribution it 
is possible to compute corrections to the semicircle law, when $N$ is 
large but finite. In particular,
it is possible to compute the correction to the averaged spectrum related  
to the  probability of having a single eigenvalue outside the semicircle 
support. 

In the context of our calculation this can be achieved
by considering solutions of the saddle point equations 
for $\rho$ different from  ${\cal S}_0$. 
In particular, we are interested in corrections to 
$\rho_0$ in the eigenvalue region on the left of 
the semicircle region, i.e. for 
$\lambda<-pE +pE_{th}$, since this tail contains the
contribution of the negative eigenvalues. 
In this region we consider the solution ${\cal S}_1$
$$
{\cal S}_1:\quad\quad\quad w_1=w_{-}(\lambda) \ \ , 
\ \ w_a=w_{+}(\lambda) \  \ ,\ \ 
a=2,\dots,n
\eq{exxola}
$$
(and permutations); from equation \re{puppa} we get 
$$
\rho_1(\lambda,E)=r(\lambda,E)\ e^{-N\Delta(\lambda,E)} \quad , 
\quad\quad \Delta(\lambda,E) > 0 \ \ {\rm for} \ \ \lambda < -pE+pE_{th}
\eq{ro1}
$$
which goes exponentially to zero as $N\to\infty$. In the computation of 
$\rho_1$  a crucial role is played by the fluctuations 
around the saddle point solution ${\cal S}_1$, since the fluctuations matrix
has an instable direction which provides  the imaginary part necessary for 
$\rho_1$ to be non-zero outside the semi-circle. On can easily check that both 
$r(\lambda,E)$ and $\Delta(\lambda,E)$ coincide with the expressions
obtained for the Gaussian random matrices with other methods \cite{metha}.
This is therefore a correct result. 
The important quantity 
for our analysis is $\Delta(\lambda,E)$
$$
\Delta(\lambda,E)={w_-^2\over 2p(p-1)}-{w_+^2\over 2p(p-1)}+
{1\over 2}\log\left(
{-\lambda-pE+w_-\over -\lambda-pE+w_+} \right)  \ .
\eq{delta}
$$ 
Solution ${\cal S}_1$ then gives  
the exponentially vanishing left tail, due to the probability of having one 
eigenvalue outside the semicircle. Since this tail is different from zero also
in the negative semi-axis, we can calculate the probability of having a 
negative eigenvalue, i.e. the exponentially small probability of finding
a TAP solution which is a saddle with one negative eigenvalue and has energy
density $E$. This probability is
$$
P_{(-)}=\int_{-\infty}^0 d\lambda \ \rho_1(\lambda,E) \sim 
\ e^{-N \Delta(0,E)}
\ \ \quad ,\quad \ \ N\to \infty
\eq{meno}
$$
In this context solution ${\cal S}_1$ has a clear physical interpretation: 
it is related to the contribution of TAP saddles 
with one negative eigenvalue, in the energy range $E<E_{th}$. Given this,
we can try to push further this interpretation. As we have seen in section 
\sref{ilcalcolo}, the same solution
${\cal S}_1$ gives rise to a complexity $\Sigma_1$ smaller than $\Sigma_0$, 
whose meaning was not clear. Now we can make the hypothesis that $\Sigma_1$
is the complexity of the saddles with one negative eigenvalue. 
To prove this statement we note that once we 
have the number ${\cal N}_1(E)\sim \exp(N\Sigma_1(E))$ of saddles with one 
negative eigenvalue and energy density $E$, we can easily 
compute the probability $P_{(-)}$ of having one of these saddles
$$
P_{(-)}={ {\cal N}_1(E) \over {\cal N}_{total}(E) }= 
{e^{N\Sigma_1(E)}\over e^{N\Sigma_0(E)}+e^{N\Sigma_1(E)}}
\sim e^{-N[\Sigma_0(E) - \Sigma_1(E)]}
\eq{selle1} 
$$
where we used the relation $ \Sigma_0(E)>\Sigma_1(E)$.
From a comparison between \re{selle1} and \re{meno} we see that it must hold
$$
\Delta(0,E)=\Sigma_0(E)-\Sigma_1(E)
\eq{pupo}
$$
It is not difficult to see from equations \re{solB}, \re{pedro} and
\re{delta} that this equation is fulfilled. Our hypothesis is
therefore correct and we can then write:
$$
\Sigma_1(E)=\lim_{N\to\infty} {1\over N} \log\overline{ {\cal N}_1(E)} 
\eq{ullamadonna}
$$
where, as already said, ${\cal N}_1(E)$ is the number of TAP solutions of
energy density $E$, which are saddles with one negative eigenvalue. 
This result can be generalized.
If we consider the following solution ${\cal S}_k$ of the saddle point 
equations for $\Sigma$
$$
{\cal S}_k:\quad q_a=q_{-} \ ,  \ a=1,\dots,k \quad  ,
\quad            q_a=q_{+} \ ,  \ a=k+1,\dots,n
\eq{nanak}
$$
(and permutations), we obtain from \re{Ba} the complexity 
$$
\Sigma_k(E)={k+2\over 2}{q_{+}^2\over p(p-1)}-{k\over 2}{q_{-}^2\over p(p-1)}
        +{k+2\over 2}\log(-pE+q_+)
        -{k\over 2}\log(-pE+q_-)
+A(E)  \ .
\eq{pedrok}
$$
It is not a surprise the fact that $\Sigma_k$ is related to the number 
of TAP solutions which are saddles with $k$ negative eigenvalues. 
Indeed the probability of finding such a solution is
$$
P_{(k,-)}=\left[ P_{(-)}\right]^k \sim e^{-N k \Delta (0,E)}
\eq{menok}
$$
so that to prove our assertion it is sufficient to verify that holds the 
relation 
$$
k\ \Delta(0,E)= \Sigma_0(E)-\Sigma_k(E)  \ 
\eq{pikappa}
$$
as it does. In writing equation \re{menok} we can disregard the correlations
between different negative ei\-gen\-va\-lues, as long as $k$ is much 
smaller than $N$.
We conclude that, as a general result, $\Sigma_k(E)$ is the complexity
of TAP saddles with $k$ negative eigenvalues and energy density $E$.

Since $|q_-|\ge|q_+|$ we have that $\Sigma_0(E)\ge\Sigma_1(E) \ge \dots 
\ge\Sigma_{k}(E)\ge \Sigma_{k+1}(E)\dots$. Thus all the TAP solutions, 
also those with some 
negative eigenvalues, are exponentially numerous in $N$. 
Nevertheless, the number of minima is exponentially higher than the 
number of saddles with one negative eigenvalue, which is exponentially higher 
than the number of  saddles with two negative eigenvalues, and so on. 
This is the very reason why, as long as $E<E_{th}$, 
the approximation of dropping the modulus in 
\re{mink} is justified. 
In Figure 1 we have plotted $\Sigma_0$, $\Sigma_1$ and $\Sigma_2$ as a 
function of $E$. 

\includegraphics{selle-fig1.ps}
\vbox{
\hbox{ \hglue   2.5 truecm \vbox{$ \Sigma$ \vglue 3.3 truecm}
\vbox{\vglue 6.7 truecm}
} \hbox{     \hglue 8 truecm $ E$ } \vbox{ \vglue 0.06 truecm} \vbox{
\hsize=16 truecm
\baselineskip=10 pt   

\piccolo{\noindent  {\pbf Figure 1}: 
The complexity $\s \Sigma_0$ of the TAP minima (solid line) and 
the complexities $\s \Sigma_1$ and  $\s \Sigma_2$ of the TAP saddles with one 
and two negative eigenvalues (respectively dotted and chain line), 
as a function of the zero temperature energy density $E$.
The threshold energy is $\s E_{th}=-1.1547$ for $\s p=3$. The minimum
saddles energy is $\s E_0=-1.1688$.
                                                           }} }
\vskip 0.5 truecm

\noindent
From equation \re{pedrok} we note that 
$\Sigma_k(E_{th})=\Sigma_0(E_{th})$, for each $k$, 
since $q_+=q_-$ at the threshold energy. This equality is very important.
If we try to count the {\it total} number of solutions neglecting the 
modulus in 
equation \re{mink}, a trivial result is obtained \cite{kurchan}, 
since we are weighting each 
stationary point with the sign of the determinant (this is the Morse theorem).
This is the reason why we considered solutions with a 
{\it given} fixed energy $E$. 
Yet, for what said above, if we integrate our result over all the
energies $E$, we must recover the result predicted by the Morse theorem.
Remembering that the $q$ part of the TAP equations admits solutions only 
for $E<E_{th}$, we have from our calculations:
$$
\eqalign{
\int^{E_{th}}  dE\ \ & 
\overline{
\int  {\cal  D} \sigma  \ \delta(\sigma \cdot \sigma -N)   
\prod_{l=1}^N\ \delta ({\cal T}_l(\sigma;E)) \  
 \det {\cal H}(\sigma;E) }= \cr   
& \phantom{
\prod_{k=1}^N\ \delta ({\cal T}_k(\sigma;E))
}
= \ a_0\ e^{N\Sigma_0(E_{th})}+ 
a_1\ e^{N\Sigma_1(E_{th})}+a_2\ e^{N\Sigma_2(E_{th})}+\dots \cr } 
\eq{muzio}
$$
In this formula we must introduce 
all the $\Sigma_k$'s coming from all the solutions 
of the saddle point equations for $\Sigma$,
which refer to stationary points of any nature. 
One can easily see that all the $\Sigma_k$'s 
are monotonously increasing functions of $E$ which reach their maximum value
at $E_{th}$, so that 
we can substitute the integral in \re{muzio} with the maximum of 
the integrand. The prefactors $a_0, a_1, \dots$ come from the 
fluctuations around each saddle point solution and contain the sign of the 
determinant.
It is exactly the combination of these signs that gives rise to the Morse 
theorem.
From equation \re{muzio} is then clear that a necessary condition to
get a trivial topological constant is that $\Sigma_0(E_{th})=\Sigma_1(E_{th})=
\Sigma_2(E_{th})=\dots$, 
so that we can sum {\it all} the terms on the same foot. 
As said above, this necessary condition is fulfilled by our calculation.

Besides, from equation \re{muzio} it is finally clear what is the role of the
modulus in the calculation: taking $\ |\det{\cal H}|\ $ is equivalent to take 
the absolute value of 
the prefactors $a_k$, thus preventing from obtaining a trivial 
result. 
Yet, at fixed energy $E<E_{th}$, one of the terms $\exp(N\Sigma_k(E))$
is always bigger than all the others and therefore in the limit $N\to\infty$
the signs of the prefactors $a_k$ have no influence on the final result.
As we have said, this dominant term turns out to be the one with $k=0$,
which gives exactly the contribution of the minima.

We note that it should be possible to show that $\Sigma_k$ is related 
to the number of the saddles with $k$ negative eigenvalue directly from
\re{Ba}. If we keep $n$ finite this integral is equivalent to 
$\overline{(\det{\cal H})^n}$. Taking the saddle point solution 
${\cal S}_k$ and appropriately computing the Gaussian fluctuations around 
it, it should be possible to 
single out a factor $(-1)^{kn}$ related to the sign of the determinant.
Unfortunately, we did not succeed in performing this quite complex 
computation.    

From figure 1 we see that there is a minimum energy
density $E_0$ below which no saddles with finite complexity are found. 
Therefore when considering a state with energy density $E<E_0$, the value
$\Delta E=E_0-E$ is a lower bound for the energy density barrier between 
this state and any other state of the system. In \cite{noi}
a potential function has been introduced, whose minima are by construction 
equivalent to metastable or equilibrium states of the system. 
With this method it has therefore been possible to give an estimate
for the barriers separating two states \cite{noi2}.
It turns out that this estimate is fully consistent with the result
of the present work. 

\autosez{conclusioni} Conclusions.
\par

\noindent
The main result of this paper concerns the organization of the
stationary points of the TAP free energy in the $p$-spin spherical model.
If we classify these points according to the number $k$ of negative 
eigenvalues of their Hessian, we find that each class is characterized
by a complexity $\Sigma_k(E)$ which gives the exponentially high number of 
TAP solutions of energy $E$ in that class, ${\cal N}_k(E)\sim 
\exp(N\Sigma_k(E))$. 
In the energy range $E<E_{th}$ we find that 
$\Sigma_k(E)>\Sigma_{k+1}(E)$ for each value of $k$. This means 
that in this energy range minima are  exponentially dominant in number 
over all the other stationary points. 

From what said above we conclude two things:
First, if we compute, even in the most rigorous way, the complexity 
$\Sigma(E)$ at a given fixed energy, according to formula \re{lei}, 
we automatically recover $\Sigma_0(E)$, i.e. the complexity of the minima. 
Secondly, the modulus of the determinant simply contributes
to the sign of the prefactor of the dominant contribution, since at
fixed energy all the other terms are vanishing in the thermodynamic
limit.
Therefore, when such a structure of the stationary points is present 
it is clear that the naive calculations which do not discriminate among
minima, saddles, etc. and which disregard the modulus are, notwithstanding 
this, consistent \cite{crisatap}\cite{noitap}.

We stress that it is crucial to keep the energy fixed in the calculation,
but more important is the fact that all the complexities are different, so that
only one of them survives in the limit $N\to\infty$. This becomes clear when 
$E$ is equal to 
the threshold energy $E_{th}$: here all the $\Sigma_k$'s are equal and a 
trivial result is recovered.

From a technical point of view we note that the use of a Bosonic 
representation for the determinant and the consequent replica approach 
introduces a degree of arbitrariness  in the choice of the saddle point 
solutions which makes it possible to extract the 
contributions of different classes of stationary points. 

\vfill \eject
\vskip 1 truecm
\noindent
{\bf References.}
\vskip .3 truecm


\biblitem{ea} Edwards S F and Anderson P W 1975 {\it J. Phys. F: Metal. Phys.} 
{\bf 5} 965

\biblitem{sk} Sherrington D and Kirkpatrick S 1975 {\it Phys. Rev. Lett.} 
{\bf 35} 1792

\biblitem{tap} Thouless D J, Anderson P W and Palmer R G 1977 {\it Philos.
Mag.} {\bf 35} 593

\biblitem{rsb1} Parisi G 1979 {\it Phys. Rev. Lett.} {\bf 23} 1754 
                          
\biblitem{rsb2} Parisi G 1980 {\it J. Phys. A: Math. Gen.} {\bf 13} L115 
                                           
\biblitem{rsb3} Parisi G 1980 {\it J. Phys. A: Math. Gen.} {\bf 13} 1887

\biblitem{sompozip} Sompolinsky H and Zippelius A 1982 {\it Phys. Rev.} B 
{\bf 25} 6860

\biblitem{ck1} Cugliandolo L F and Kurchan J 1993 {\it Phys. Rev. Lett.}
{\bf 71} 173

\biblitem{tirumma} Kirkpatrick T R and Thirumalai D 1987 {\it Phis. Rev.} B 
{\bf 36} 5388

\biblitem{crisa1} Crisanti A and Sommers H-J 1992 {\it Z. Phys.} B 
{\bf 87} 341

\biblitem{crisa2} Crisanti A, Horner H and Sommers H-J 1993 {\it Z. Phys.} B
{\bf 92} 257

\biblitem{ck2} Cugliandolo L F and  Kurchan J 1994 {\it J. Phys. A: Math. Gen.} 
{\bf 27} 5749

\biblitem{kpz} Kurchan J, Parisi G and Virasoro M A 1993 {\it J. Phys. I France}
{\bf 3} 1819

\biblitem{franzparisi} Franz S and Parisi G 1995 {\it J. Phys. I France}
{\bf 5} 1401

\biblitem{ferrero} Ferrero M E and Virasoro M A 1994 {\it J. Phys. I France }
{\bf 4} 1819

\biblitem{buribarrameza} Barrat A, Burioni R and M\'ezard M 1996 
{\it J. Phys. A: Math. Gen.}
{\bf 29} L81

\biblitem{monasson} Monasson R 1995 {\it Phys. Rev. Lett.}
{\bf 75} 2847

\biblitem{crisatap} Crisanti A and Sommers H-J 1995 {\it J. Phys. I France}
{\bf 5} 805

\biblitem{mezpa} M\'ezard M and Parisi G 1990 {\it J. Phys. A: Math. Gen.}
{\bf 23} L1229; M\'ezard M and Parisi G 1991 {\it J. Phys. I France } 
{\bf 1}  809

\biblitem{nieu} Nieuwenhuizen Th M 1996 {\it Phys. Rev. Lett. }
{\bf 74} 4289

\biblitem{franz} Franz S, private communication

\biblitem{vira} Virasoro M A {\sl Simulated annealing methods under analytical 
control}, in Pro\-cee\-dings of 19th Intl. Conf. on Stat. Phys. (IUPAP),
Xiamen, China; to be published

\biblitem{grome} Gross D J and M\'ezard M 1984 {\it Nucl. Phys. B }
{\bf 240} 431

\biblitem{gard} Gardner E {\it Nucl. Phys. B}
{\bf 257} 747

\biblitem{noi} Cavagna A, Giardina I and Parisi G 1997 {\it J. Phys. A: Math. 
Gen.} {\bf 30} 4449

\biblitem{noitap} Cavagna A, Giardina I and Parisi G 1997 {\it J. Phys. A: 
Math. Gen.} {\bf 30} 7021

\biblitem{noi2} Cavagna A, Giardina I and Parisi G 1997 cond-mat 9702069

\biblitem{braymoore} Bray A J and Moore M A 1980 {\it J. Phys. C: Solid. St. 
Phys.} {\bf 13} L469

\biblitem{frapavi} Franz S, Parisi G and Virasoro M A 1992 {\it J. Phys. I
France} {\bf 2} 1869
 
\biblitem{kurchan} Kurchan J 1991 {\it J. Phys. A: Math. Gen.}
{\bf 24} 4969

\biblitem{vv} Vertechi D and Virasoro M A 1989 {\it J. Physique}
{\bf 50} 2325

\biblitem{bfp} Barrat A, Franz S and Parisi G cond-mat 9703091

\biblitem{rem} Derrida B 1981 {\it Phys. Rev.} {\bf B 24} 2613

\biblitem{ultra} M\'ezard M, Parisi G, Sourlas N, Toulouse G and Virasoro M A
1984 {\it J. Physique} {\bf 45} 843

\biblitem{spin} M\'ezard M, Parisi G and Virasoro M A 1986 {\sl Spin Glass 
Theory And Beyond}, World Scientific Pu\-bli\-shing, Singapore

\biblitem{metha} Metha M L 1967 {\it Random Matrices} (Academic Press, 
New York)

\biblitem{wigner} Wigner E P 1957 Can. Math. Congr. Proc. (Univ. of Toronto
Press, Toronto), reprinted in Porter C E 1965 {\it Statistical theories
of spectra: fluctuations} (Academic Press, New York)

\biblitem{morse} Doubrovine B, Novikov S and Fomenko A 1982 
{\it Geometrie Contemporaine} vol 2, ch 14 (Moscow: Mir) 

\insertbibliografia

\bye